\documentclass[aps, prb, reprint, amssymb, amsfonts, superscriptaddress,longbibliography ]{revtex4-1}

\usepackage{amsmath}
\usepackage{bm}
\usepackage{graphicx}
\usepackage[unicode]{hyperref}
\usepackage[latin1]{inputenc}
\usepackage{array}

\begin{document}

\title{Probing the Nuclear Spin-lattice Relaxation time at the Nanoscale}

\author{J. J. T. Wagenaar}
	\email{wagenaar@physics.leidenuniv.nl}
\author{A. M. J. den Haan}
\author{J. M. de Voogd}
\author{L. Bossoni}
\author{T. A. de Jong}
\author{M. de Wit}
\author{K. M. Bastiaans}
	\affiliation{Kamerlingh Onnes Laboratory, Leiden University, P.O. Box 9504, 2300 RA Leiden, The Netherlands}

\author{D. J. Thoen}
	\affiliation{Kavli Institute of Nanoscience, Faculty of Applied Sciences, Delft University of Technology, Lorentzweg 1, 2628 CJ Delft, The Netherlands} 
\author{A. Endo}
	\affiliation{Kavli Institute of Nanoscience, Faculty of Applied Sciences, Delft University of Technology, Lorentzweg 1, 2628 CJ Delft, The Netherlands} 
	\affiliation{Department of Microelectronics, Faculty of Electrical Engineering, Mathematics and Computer Science, Delft University of Technology, Mekelweg 4, 2628 CD Delft,The Netherlands} 
\author{T. M. Klapwijk}
	\affiliation{Kavli Institute of Nanoscience, Faculty of Applied Sciences, Delft University of Technology, Lorentzweg 1, 2628 CJ Delft, The Netherlands} 
	\affiliation{Moscow State Pedagogical University, 1 Malaya Pirogovskaya Street, Moscow 119992, Russia} 

\author{J. Zaanen}
	\affiliation{Instituut-Lorentz for Theoretical Physics, Leiden University, P.O. Box 9506, 2300 RA Leiden, The Netherlands}

\author{T.H. Oosterkamp}
\affiliation{Kamerlingh Onnes Laboratory, Leiden University, P.O. Box 9504, 2300 RA Leiden, The Netherlands}

\date{\today}

\begin{abstract}
Nuclear spin-lattice relaxation times are measured on copper using magnetic-resonance force microscopy performed at temperatures down to $42$ mK. The low temperature is verified by comparison with the Korringa relation. Measuring spin-lattice relaxation times locally at very low temperatures opens up the possibility  to measure the magnetic properties of inhomogeneous electron systems realized in oxide interfaces, topological insulators and other strongly correlated electron systems such as high-T$_c$ superconductors.
\end{abstract}

\maketitle

\section{Introduction}

Among the most informative probes of  electron systems in solids is the nuclear spin-lattice relaxation rate $1/T_1$. This value quantifies the damping of the nuclear spin precession due to the coupling to the electron spins. This in turn can be related to the momentum-averaged imaginary part of the dynamical spin susceptibility of the electron system, measured at the very low Larmor frequency of the nuclear spins. This is a most useful quantity revealing basic physics of the electron system. For instance, in Fermi liquids like copper, the Korringa relation \cite{Korringa1950}, $1/(T_1T) =$~constant, universally holds, while non-Korringa behaviors play a pivotal role in establishing the unconventional nature of various strongly interacting electron systems \cite{Alloul2014}. This mainstay of traditional NMR methods is hampered by the fact that it is a very weak signal that usually can  be detected only in bulk samples \cite{Glover2002}.  We demonstrate here an important step forward towards turning this into a nanoscopic probe, by delivering a proof of principle that $1/T_1$ can be measured at subkelvin temperatures using magnetic-resonance force microscopy (MRFM). State-of-the-art MRFM demonstrates an imaging resolution of $<10$ nm, a 100-million-fold improvement in volume resolution over bulk NMR, by detecting the proton spins in a virus particle \cite{Degen2009}. In this experiment the statistical polarization of the protons is measured, since the Boltzmann polarization is too small to detect. 

Measurements of the recovery of the Boltzmann polarization after radio-frequent pulses relate directly to spin-lattice relaxation times. The force-gradient detection of $T_1$ is measured before in GaAs at temperatures down to $4.8$ K \cite{Alexson2012}.  In this paper we demonstrate $T_1$ measurements [see Fig. \ref{fig:figure3}(a)] at temperatures which are lower by 2 orders of magnitude and a volume sensitivity of 3 orders of magnitude larger. Furthermore, we obtain the temperature dependence of the relaxation time $T_1$ satisfying quantitatively the Korringa relation  of copper; see Fig. \ref{fig:figure4}. This is a major step forward towards the ultimate goal of a device that can measure  variations of $T_1$ at the nanoscale to study the properties of the inhomogeneous electron systems which are at the forefront of modern condensed-matter physics, such as the surface states of topological insulators \cite{Chen2009}, oxide interfaces \cite{Mannhart2010, Richter2013, Kalisky2013, Scopigno2016} and other strongly correlated electron systems \cite{Dagotto2005} including the high-T$_c$ superconductors \cite{Keimer2015}.    

\begin{figure}[hbt] 
	\centering
		\includegraphics[width=\columnwidth]{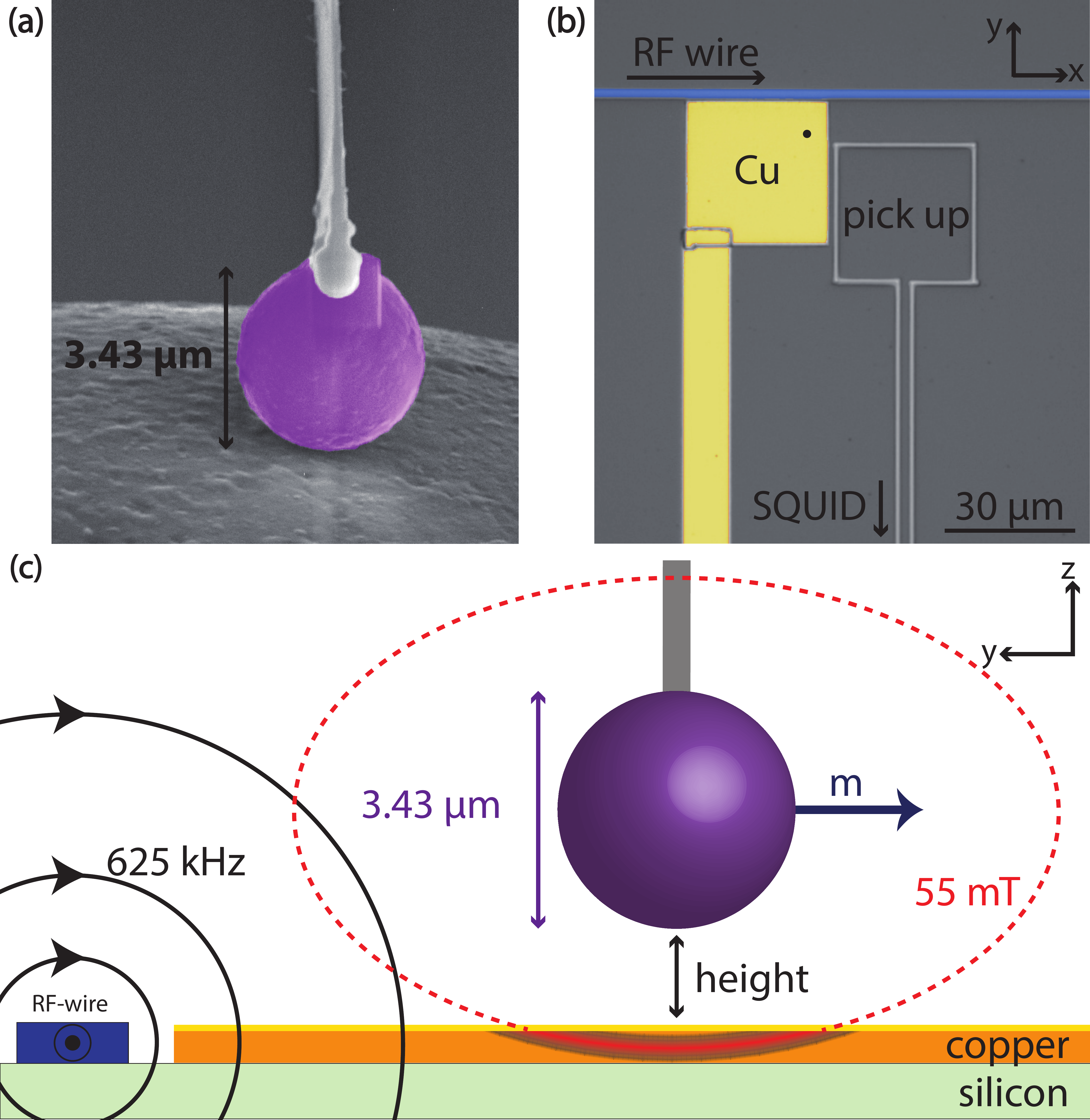}
		\caption{\textbf{a}) Scanning-electron-microscope image (false color) of the magnetic particle after it is glued to the cantilever using electron-beam-induced deposition from a platinum-containing precursor gas.  \textbf{b}) Optical-microscope image (false colors) of the detection chip. The blue horizontal wire at the top is the NbTiN rf wire . The gold-capped copper sample is connected to the mixing chamber for thermalization. The small black dot is where we perform the MRFM experiments. Next to the sample and below the rf wire is the pickup coil for detection of the cantilever's motion. \textbf{c}) Schematic drawing of a side view of the local NMR experiment. When the rf frequency meets the resonance condition $f_{rf}=\frac{\gamma}{2\pi}B_0$, copper nuclei at magnetic-field strength $B_0$ are in resonance. }
	\label{fig:figure1}
\end{figure}

MRFM uses a magnetic particle attached to an ultrasoft cantilever [Fig. \ref{fig:figure1}(a)] to detect the force gradients between sample spins and the tip. In most MRFM apparatuses a laser is used to read out the motion of the cantilever. The spins in the sample are manipulated using radio-frequent (rf) currents that are applied using a copper or gold nanowire \cite{Poggio2007, Isaac2016}. The laser and large rf currents in the copper or gold nanowires prevent experiments at millikelvin temperatures. To our knowledge, sample temperatures below $4.2$~K have not been unequivocally demonstrated, only that the working temperature of the cryostat can remain at $300$~mK \cite{Degen2009}. In our setup, a superconducting quantum-interference device (SQUID) \cite{Usenko2011, Vinante2011, Vinante2012} is used combined with a superconducting rf wire [Fig. \ref{fig:figure1}(b)]. This enables us to perform  nuclear MRFM experiments down to temperatures of $42$ mK, and we are able to verify that the sample indeed is at this temperature.

\section{Methods}

\subsection{Experimental setup}

The experimental setup closely resembles the one that was recently used to measure  the dangling bonds on a  silicon substrate \cite{Haan2015}. The NdFeB magnetic particle is $3.43\pm0.02$ $\mu$m in diameter, and has a saturation magnetization of $1.15$ T. The cantilever's natural resonance frequency is $f_0=3.0$ kHz with a mechanical quality factor of $Q_0=2.8\cdot10^4$, when far away from the surface. An in-house-developed cryopositioning system is used to move the cantilever relative to the sample  with  a range of $1$ mm in all dimensions. The absolute position of the cantilever is measured using three capacitive sensors. The sample holder is placed on a fine stage which is used to change the height of the sample in a range of $2.3$ $\mu$m. 

For the fabrication of the detection chip consisting of a superconducting rf wire, a pickup coil, and the copper sample, we start with a NbTiN film on a silicon wafer with its natural oxide. The film has an average thickness of $378$ nm, a resistivity of $92$ $\mu \Omega$~cm and a critical temperature of $15.3$ K. Using standard electron-beam-lithography techniques, with a negative resist, we pattern the rf wire structure and pickup coil.  We use reactive ion etching in a SF$_6$/O$_2$ plasma to etch the film. The copper is sputtered with a thickness of $300$ nm and a roughness of $10$ nm. In order to prevent oxidation of the copper, a gold capping layer of $15$ nm is sputtered while the sample remains in high vacuum. The copper sample is thermalized using a patterned copper wire leading to a large copper area, which is connected to the sample holder via gold-wire bonds. The sample holder is connected to the mixing chamber of the dilution refrigerator via a welded silver wire.

We position the cantilever above the copper, as indicated by a small black dot in Fig. \ref{fig:figure1}(b), close ($7\pm1$~$\mu$m) to the center of the rf wire and close ($5\pm1$~$\mu$m) to the pickup coil for sufficient signal strength. While approaching the copper, we measure an enormous drop in the quality factor of the cantilever towards $Q=300$ which strongly depends on the distance to the sample. This is caused by eddy currents in the copper induced by the magnetic fields of the magnetic particle attached to the moving cantilever. The drop in the quality factor limits the minimal height above the surface, reducing the signal strength. 

Natural copper exists in two isotopes. Both isotopes have a spin $S=3/2$, $^{63}$Cu has a natural abundance of $69$\% and a gyromagnetic ratio $^{63}\gamma/(2\pi)=11.3$ MHz/T and $^{65}$Cu has a natural abundance of $31$\% and a gyromagnetic ratio $^{65}\gamma/(2\pi)=12.1$ MHz/T. In fcc copper, there is no quadrupolar coupling, since electric-field gradients are zero given the cubic symmetry.

The magnetic particle on the cantilever couples to all nuclear spins in the sample \textit{via} its field gradient. Every spin effectively shifts  the resonance frequency of the cantilever by an amount $\Delta f_s$. The alignment of the nuclear spins is perturbed when a radio-frequent pulse is applied to the sample. The spins that meet the resonance condition $f_{rf}=\frac{\gamma}{2\pi} B_0$ are said to be in the resonant-slice, where $f_{rf}$ is the rf frequency and $B_0$ the local magnetic field.  The total frequency shift due to all the spins within a resonant-slice can be calculated by the summation of the single-spin contributions $\Delta f_0=p\sum_s \Delta f_s$, where $p$ is the net Boltzmann polarization of the nuclear spins. Those spins losing their net Boltzmann polarization (i.e.,  $p\rightarrow0$)  will cause a total frequency shift of $-\Delta f_0$. This effect can be obtained at low radio-frequent magnetic fields in a saturation experiment with low enough currents to prevent heating of the sample. 

A rf current of $0.2$ mA is applied in all measurements. Taking the approximation of a circular wire, we estimate the typical field strength in the rotating frame for a current of $0.2$ mA at a distance $r=7\ \mu$m to be $B_1=\frac{\mu_0 I}{4\pi r}\approx 2.9\ \mu$T. The rf-field is mostly perpendicular to $B_0$ (see Fig. \ref{fig:figure1}(c)). The saturation parameter  $s\equiv\gamma^2 B_1^2 T_1 T_2$ \cite{Abragam1961} is used to calculate the ratio of total saturation which is given by $\frac{s}{1+s}$. For temperatures $T<250$ mK, the expected values for the relaxation times are $T_1>5$ s. Given a  $T_2=0.15$ ms \cite{Pobell2007}, the saturation parameter becomes $s>30$, indicating that the spin saturation is at least $95$\%,  for those spins that satisfy the resonance condition. Since we work with field gradients, there will always be spins which are not fully saturated. For the sake of simplicity, we assume that we have a resonant-slice thickness $d$, within which the spins are fully saturated. The Rabi frequency is $\frac{\gamma}{2\pi} B_1\approx30$ Hz, so we assume that for pulse lengths of $1$ s all levels are fully saturated.  When all levels saturate, the magnetization after a pulse restores according to a single exponential, with a decay time equal to the spin-lattice relaxation time $T_1$. The frequency shift of the cantilever is proportional to the magnetization, and  the time-dependent resonance frequency $f(t)$ becomes:
\begin{align}
f(t)=f_0+\Delta f_0 \cdot e^{-(t-t_0)/T_1}
\label{eq:fitting}
\end{align}
We use the phase-locked loop (PLL) of a Zurich Instruments lock-in amplifier to measure the shifts in resonance frequency $\delta f=f(t)-f_0$ of the cantilever at a bandwidth of $40$ Hz. The measurement scheme is as follows: First we measure the natural resonance frequency $f_0$ using a PLL. The PLL is turned off, and the rf current is turned on for 1 s. At $t_0$, the rf current is turned off. Shortly thereafter, the PLL is turned on, and we measure the frequency shift relative to $f_0$. The PLL is switched off during and shortly after the rf pulse in order to avoid cross talk.

At frequencies larger than $1$ MHz and currents higher than $1$ mA, we measure about a few millikelvins increase in the temperature of the sample. The dissipation of the rf wire will be subject of further study, since for nanoscale imaging based on (proton) density, the magnetic rf fields need to be large ($>3$ mT) enough for adiabatic rapid passages \cite{Poggio2007}. In this paper, we avoid the large currents in order to prevent heating.

\begin{figure}[t]
	\centering
		\includegraphics[width=\columnwidth]{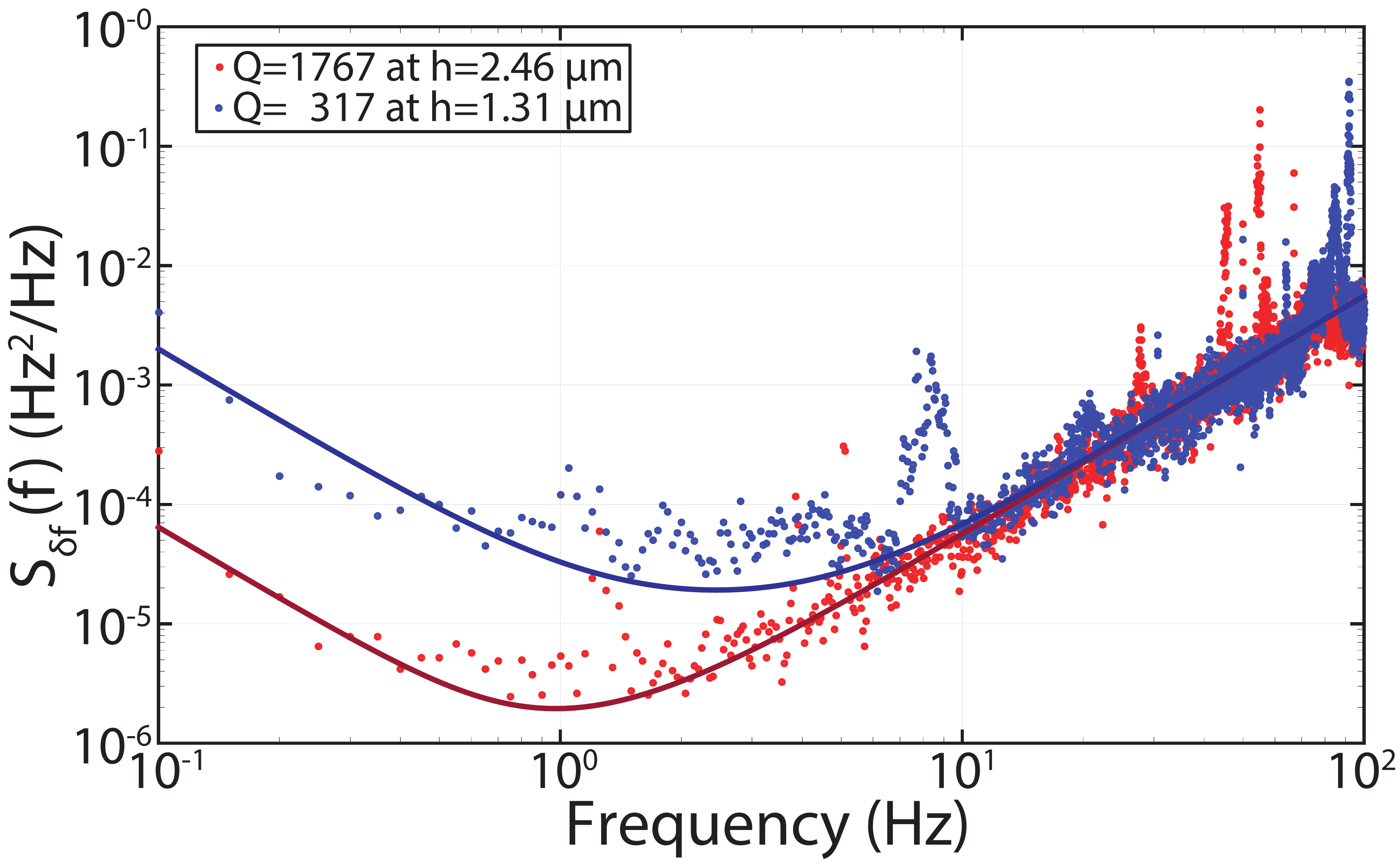}
		\caption{The frequency noise $S_{\delta f}(f)$ for two different cantilever heights above the copper sample. The solid line is the expected frequency noise according to Eq. \ref{eq:freqnoise} with an added $1$/f noise. The Q dependence of the 1/f noise suggests that the additional noise is created by the oscillating currents in the copper, which also cause the drop in the quality factor due to eddy currents. }
	\label{fig:figure2}
\end{figure}

\subsection{Frequency noise}
We determine the shifts in resonance frequency $\delta f$ using a phase-ocked loop. The noise spectrum of the frequency shifts $S_{\delta f}(f)$, which has units Hz$^2$/Hz, consists of the thermal noise of the cantilever and detector noise $S_{x_{det}}(f)$. References \cite{Albrecht1991,Kobayashi2009, Isaac2016} derive the frequency noise for frequency-modulation detection. We obtain the following expression, without the commonly used approximation $f\gg\frac{f_0}{2Q}$, and with $A$ the rms amplitude for the cantilever displacement:
\begin{align}
S_{\delta f}(f)=\frac{4k_BTQ}{2 \pi A^2 k_0 f_0}\frac{f^2}{1+\left(\frac{2Qf}{f_0}\right)^2}+\frac{S_{x_{det}}(f)}{A^2}\left(f^2+\frac{f_0^2}{4Q^2}\right)
\label{eq:freqnoise}
\end{align}
The detector noise $S_{x_{det}}(f)$ is determined by measuring the voltage noise $S_{V_{det}}(f)$ of the SQUID. The voltage noise corresponds to a certain position noise of the cantilever. By measuring the thermal spectrum of the cantilever motion at different temperatures and by using the equipartition theorem $\left<x^2\right>=\frac{k_BT}{k_0}$, we calibrated the transfer between the position and the SQUID's output voltage. With this, we obtain $\sqrt{S_{x_{det}}(f)}=82$ pm/$\sqrt{\text{Hz}}$. The thermodynamic temperature of the cantilever is determined to be $139$ mK and the cantilever motion is kept constant at a rms amplitude of $A=110$ nm. Representative measurements for two different heights are shown in Fig. \ref{fig:figure2}. 

The solid line in Fig. \ref{fig:figure2} is Eq. (\ref{eq:freqnoise}) with no fitting parameters but with an added noise source equal to $S_{add}(f)=\frac{1 \text{Hz}^3}{Q^2 f^2}$. The numerical factor of $1$ is  chosen just for convenience, but the dependence on $Q$ indicates that the additional noise is created by the oscillating currents in the copper, which also cause the drop in the quality factor due to eddy currents. In other MRFM apparatuses, the 1/f noise is shown to be caused by dielectric fluctuations \cite{Isaac2016, Kuehn2006}, but in those setups the tip-sample distance is much smaller than in our case. We hope that nuclear MRFM on other systems than copper result in a smaller drop in the quality factor and therefore in a much better signal-to-noise ratio.

\subsection{Numerical calculation of frequency shifts}
To calculate the frequency shifts $\Delta f_0$ due to the spins in resonance [Eq. \ref{eq:fitting}], the contributions of all spins meeting the resonance condition are integrated. Traditionally, the measured frequency shift of a mechanical resonator in MRFM is simulated using $k=-\left<m_z\right> B''_z$, the stiffness shift due to the coupling to a single spin \cite{Isaac2016}. However, a recent theoretical analysis by De Voogd, Wagenaar, and Oosterkamp \cite{Voogd2015},  supported by experiments performed by Den Haan \textit{et al.}\cite{Haan2015}, suggests that this is only an approximation. Following the new analysis, using $f_{rf}\gg \frac{1}{T_2}\approx 6$ kHz, we obtain for the single-spin contribution $\Delta f_s$
\begin{align}
\Delta f_s&=\frac{1}{2}\frac{f_0}{k_0}Re{(\left(F_1+F_2+F_{34}\right))}\\
F_1&=-\left<m\right>|\bm{B''_{||B_0}}|\\
F_2&=-\frac{\left<m\right>'}{B'_0}|\bm{B'_{||B_0}}|^2\frac{1}{1+i\omega T_1}\\
F_{34}&=- \frac{\left<m\right>}{B_0} |\bm{B'_{\perp B_0}}|^2
\end{align}
The primes and double primes refer to the first and second derivative, respectively, with respect to the fundamental direction of motion of the cantilever $x$, which is in our experimental setup the $x$ direction. $|\bm{B''_{||B_0}}|$ is the component along $B_0$.  $|\bm{B'_{\perp B_0}}|$ is the perpendicular component . The term $F_2$ vanishes because $\omega T_1\gg 1$. $\left<m\right>$ is the mean (Boltzmann) polarization in units J/T. 

For arbitrary spin $S$ we find for $\left<m\right>$ 
\begin{align}
 \left<m\right> &=   \gamma \hbar  \left[ \left(S+\frac{1}{2}\right)
 \coth{\left(\frac{\gamma \hbar B_0\left(S+\frac{1}{2}\right)}{k_BT}\right)}       \right.\nonumber\\
   &\qquad \qquad \qquad \qquad \qquad -\frac{1}{2}\coth{\left(\frac{\gamma \hbar B_0}{2k_BT}\right)} \Bigg ].
\end{align}
In order to find the total direct frequency shift  $\Delta f_0$ of a saturated resonant-slice with thickness $d$, we need to integrate all terms over the surface of the resonant-slice:
\begin{align}
\Delta f_0=\rho d \iint_{slice} \Delta f_s dS
\end{align}
with $\rho$ the spin density. To simplify the integration, we switch to spherical coordinates, with $r$ the radius from the center of the magnet to the position in the sample, $\phi$ the angle running from $0$ to $2\pi$ within the plane and $\theta$ the angle with the \textit{z} axis, which is perpendicular to the copper surface. When the magnet is centered above a uniform sample that extends sufficiently far in the plane, we can assume that $\phi$ can be integrated from $0$ to $2\pi$. The distance to the resonant-slice $r_{res}$ can be expressed as a function of $B_0$, $\theta$ and $\phi$.  The upper boundary of $\theta$ equals $\pi$ for an infinitely thick sample. The lower boundary $\theta_{min}$ can be expressed analytically using a geometric relation and the expression for the magnetic field of a magnetic dipole $M$: 
\begin{align}
h&=r_{res}(\theta_{min},\phi)\cos{\theta_{min}}\\
B_0&=\frac{\mu_0 M}{4\pi r^3}\sqrt{3\sin^2{\theta}\sin^2{\phi}+1}
\end{align}
Combining the above equations we obtain an expression for $\theta_{min}$:
\begin{align}
0=&\left(3\sin^2{\theta_{min}}\sin^2{\phi}+1\right)\cos^6{\theta_{min}}-\left(\frac{4\pi B_0h^3}{\mu_0M}\right)^2
\end{align}
Solving this  equation gives us the final unknown integration boundary $\theta_{min}$. 
\begin{align}
\Delta f_0=\rho d \int_{\theta_{min}}^{\pi}d\theta \int_0^{2\pi} \Delta f_s(\theta, \phi, B_0, d) d\phi
\end{align}
The signal coming from a sample of thickness $d_{Cu}$ at height $h$ is equal to an infinite sample at height $h$ minus an infinite sample at distance $h+d_{Cu}$. All parameters relevant for the experiment on copper are listed in Table \ref{table:variables}. The only free parameters are the height above the surface $h$, the absolute value of which we are not able to determine accurately because of the eddy currents close at small heights, and the thickness of the resonant-slice.

\begin{table}[tb]
\centering
\caption{Overview of the setup parameters and NMR constants and parameters for copper \cite{Lounasmaa1974, Huiku1986, Oja1997, Pobell2007}.}
\label{table:variables}
\begin{tabular}{ccc}
\hline\hline
Parameter                    & Variable      & Value          \\	 
\hline
Stiffness cantilever		 & $k_0$		 & $7.0\cdot10^{-5}$ Nm$^{-1}$\\
Bare resonance frequency	 & $f_0$		 & $3.0$ kHz      	\\
Intrinsic quality factor	 & $Q_0$		 & $2.8\cdot10^4$	\\
Radius magnet                & $R_0$         & $3.43/2$ $\mu$m 	\\
Magnetic dipole              & $M$           & $1.9\cdot10^{-11}$ Am$^{2}$        	\\
Copper-layer thickness       & $d_{Cu}$      & $300$ nm        	\\
Copper spin density			 & $\rho$		 & $85$ nm$^{-3}$	\\
Spin copper nuclei			 & $S$			 & $3/2$			\\
Gyromagnetic ratio $^{63}$Cu & $^{63}\gamma/(2\pi)$ & $11.3$ MHz/T    	\\
Gyromagnetic ratio $^{65}$Cu & $^{65}\gamma/(2\pi)$ & $12.1$ MHz/T    	\\
Natural abundance $^{63}$Cu  &               & $69$ \%         	\\
Natural abundance $^{65}$Cu  &               & $31$ \%         	\\
Korringa constant $^{63}$Cu	 & $\kappa_{63}$ & $1.27$ sK     	\\
Korringa constant $^{65}$Cu  & $\kappa_{65}$ & $1.09$ sK     	\\
Spin-spin relaxation time    & $T_2$         & $0.15$ ms     	\\
\hline\hline 
\end{tabular}
\end{table}

\begin{figure}[tb]
	\centering
		\includegraphics[width=\columnwidth]{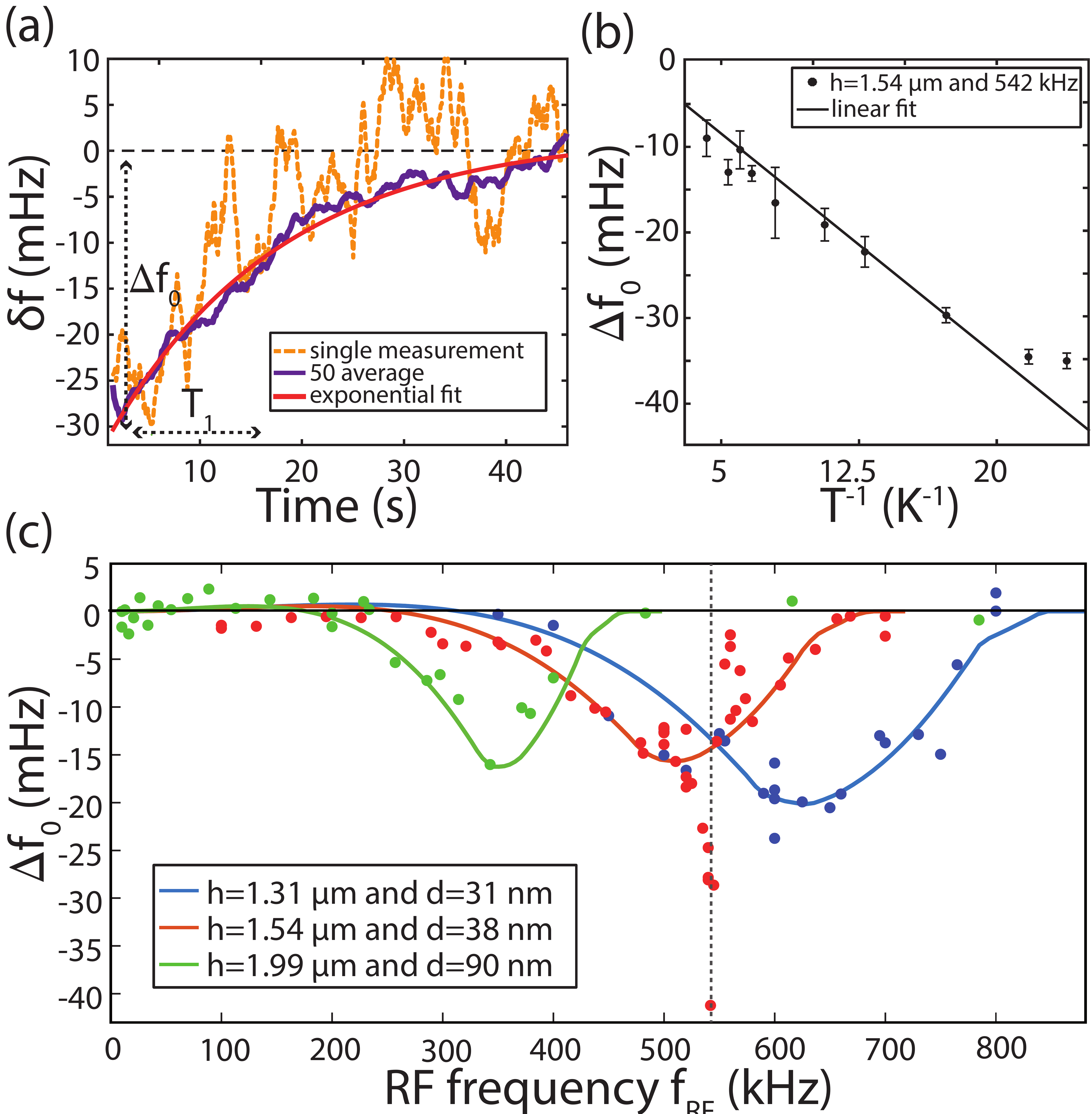}
		\caption{\textbf{a}) The saturating rf pulse results in a direct frequency shift $\Delta f_0$ of the cantilever's resonance frequency. The shift relaxes with the nuclear spin's relaxation time $T_1$. The dashed orange line shows a single measurement of the frequency shift (with a moving average of 1 s); the solid blue line shows $50$ averages. The red solid line is an exponential fit according to equation \ref{eq:fitting}. From the fit we extract the direct shift $\Delta f_0$, the relaxation time $T_1$, and a residual frequency offset $\delta f_0$. \textbf{b}) The direct frequency shift $\Delta f_0$ versus the temperature. The linear fit is expected according to Curie's law $\langle m \rangle\propto T^{-1}$, which implicitly assumes that the resonant-slice thickness is temperature independent. \textbf{c}) $\Delta f_0$ as function of rf-frequency $f_{rf}$ for three different cantilever heights at $T=42$ mK. The solid line is the calculated signal with the cantilever height $h$ and resonant-slice width $d$ as parameters. The two resonant-slices of the two copper isotopes largely overlap each other. The dashed vertical line at $542$ kHz indicates the location of a higher resonance mode, at which the cantilever can be driven using the rf-frequency resulting in an amplification of the $B_1$ field and, therefore, the spin signal (see the main text).}
\label{fig:figure3}
\end{figure}

\section{Results and discussion}

\subsection{Direct frequency shifts}
In Fig. \ref{fig:figure3}(a) we show a representative plot of a saturation recovery measurement performed at a temperature of $42$ mK and a rf frequency of $542$ kHz. The cantilever is positioned at a height of approximately $1.5$ $\mu$m above the surface of the copper.  At $f_{rf}=542$ kHz, the measured frequency shifts are larger than those observed at other frequencies. This is caused by the presence of a higher resonance mode of the cantilever which can be driven by the magnetic field of the rf wire. The resulting increased motion of the magnetic particle at the end of the cantilever gives an additional rf-field that will increase the number of spins that are saturated. The exact mechanism for this will be presented in a separate paper \footnote{A. M. J. den Haan, J. J. T. Wagenaar, and T. H. Oosterkamp, A Magnetic Resonance Force Detection Apparatus and Associated Methods, United Kingdom Patent No. GB 1603539.6 (1 Mar 2016), patent pending}.  We measure the frequency shift up to $t= 50$ s and average $50$ times. Subsequently, the data are fitted to Eq. (\ref{eq:fitting}) to extract $\Delta f_0$, $T_1$ and a residual frequency offset $\delta f_0$.

The direct frequency shift $\Delta f_0$ is plotted versus the temperature in Fig. \ref{fig:figure3}(b) and fitted with a straight line according to the expected Curie's law $\Delta f_0\propto \langle m \rangle \propto \frac{1}{T}$, which implicitly assumes that the resonant-slice thickness is temperature independent. $\Delta f_0$ shows a saturation at the lowest temperatures. This indicates that the (electron) spins are difficult to cool, although the saturation could also be caused by the approach of the minimum temperature of the refrigerator.  For every temperature, we collect at least three sets of $50$ curves. The error bars are the standard deviations calculated from the separate fits. 

In Fig. \ref{fig:figure3}(c) the direct frequency shift is plotted versus rf-frequency for three different cantilever heights. Every data point resembles a data set of $50$ averaged single measurements. The solid line is a numerical calculation, with the only fitting parameters the resonant-slice width and height of the cantilever above the surface. The height of the cantilever is left as a fitting parameter, because the absolute height is not known with sufficient accuracy, due to the nonlinear behavior of the piezostack in the fine stage.

Several mechanisms will contribute to the resonant-slice thickness; for example, the width of the resonant-slice is determined by the NMR linewidth, i.e., the saturation parameter, and may be further broadened by the nuclear spin diffusion and by the resonant displacement of the cantilever.  These various contribution are discussed in more detail below. 

\subsection{The Korringa relation}

\begin{figure}[tb]
	\centering
		\includegraphics[width=\columnwidth]{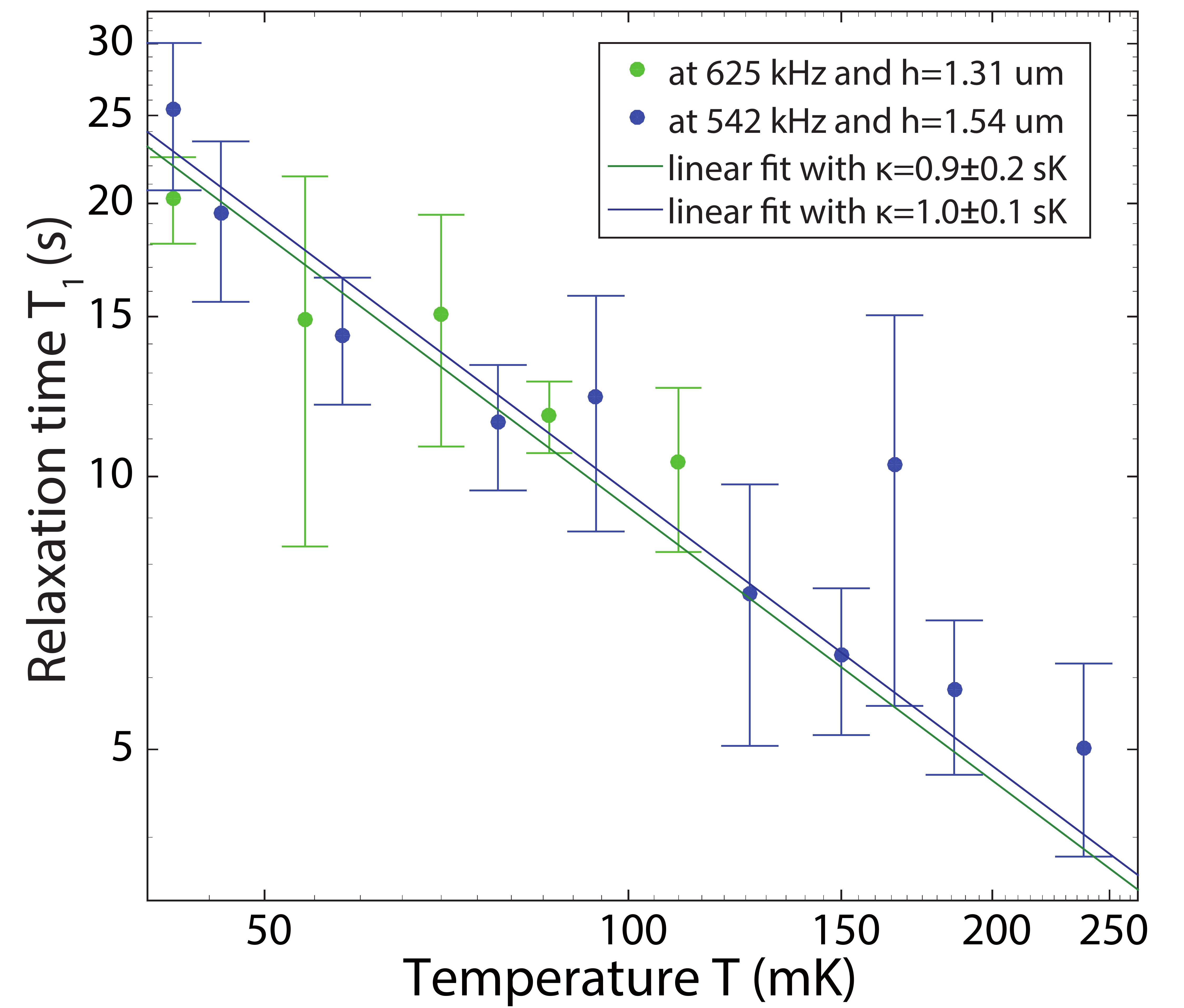}
		\caption{The inverse of the relaxation time of the copper nuclei measured by applying saturation pulses versus the temperature. We find the same linear dependence with temperature as is found in bulk copper samples, the so-called Korringa relation. In blue are data measured at a height of $1.54$ $\mu$m at $542$ kHz, where a larger signal is achieved due to amplification of the rf-field by a higher cantilever mode. In green are data recorded at $1.31$ $\mu$m at a frequency of $625$ kHz. We find $\kappa=1.0\pm0.1$ sK and $\kappa=0.9\pm0.2$ sK, respectively, with a linear fit with the error indicating the 95\% confidence intervals. Every data point is an average of at least three sets of averaged data. The error bars give the standard deviation of the found relaxation times for the averaged data sets.  }
	\label{fig:figure4}
\end{figure}

For two different rf-frequencies, $542$ and $625$ kHz, at two different heights, $1.54$ and $1.31$ $\mu$m, we measure the relaxation time $T_1$ versus the temperature. The results are shown in Fig. \ref{fig:figure4}. We observe a linear dependence of $T_1$ on the temperature, following the Korringa law $T_1T=\kappa$. $T$ is the electron temperature, and $\kappa$ is the Korringa constant. From the linear fits, we extract $\kappa=1.0\pm0.1$ sK and $\kappa=0.9\pm0.2$ sK, which is close to the expected value of the combined $^{63}$Cu and $^{65}$Cu Korringa constants, $\kappa=1.2$ sK \cite{Pobell2007}, which is measured in bulk copper.

\subsection{The resonant-slice thickness}
The thickness of the resonant-slice is important for the resolution of possible future imaging or spectroscopic experiments or when similar measurements are to be performed on much thinner films. We also need the resonant-slice width when modeling the signals we measured in Fig. \ref{fig:figure3}(c), because they are determined by the number of spins taking part. This results in a fit parameter of the resonant-slice thicknesses of $90$, $38$ and $31$ nm (for heights of $1.31$, $1.54$, and $1.99$ $\mu$m, respectively). In the case of our experiment on copper, we believe that the width of the resonant-slice is determined by the NMR linewidth, i.e., the saturation parameter, and may be further broadened by the nuclear spin diffusion and by the resonant displacement of the cantilever. Below, we discuss these various contributions.

First, consider the steady-state solution for the $z$ component of the magnetization $m_z$ of the spins in a conventional magnetic-resonance saturation experiment \cite{Abragam1961}:
\begin{equation}
m_z=\frac{1+(2\pi T_2\Delta f )^2}{1+(2\pi T_2\Delta f)^2+\gamma^2 B_1^2 T_1 T_2} \langle m \rangle,
\end{equation}
where $\Delta f$ is the detuning of the rf frequency from the Larmor frequency and $s=\gamma^2 B_1^2 T_1 T_2 $ is the saturation parameter. $\langle m \rangle$ is the magnetization in equilibrium. 

The value $\Delta f$ for which $m_z=\frac{1}{2}\langle m \rangle$ is a measure for the resonance thickness by $d=\frac{4\pi \Delta f}{\gamma \nabla_r B_0}$ with $\nabla_r B_0$ the field gradient in the radial direction. We find a thickness of $98$, $58$ and $45$ nm for, respectively, the fitted thicknesses of $90$, $38$, and $31$ nm.

One additional possibility to broaden the slice in the radial direction is the effect of nuclear diffusion. For polycrystalline copper, we find a transition rate of $W=\frac{1}{30T_2}$ \cite{Abragam1961,Bloembergen1949}. Using $T_2=0.15$ ms, we find $W=2.2\cdot10^2$~s$^{-1}$. Our field gradients are too small to quench the spin-diffusion \cite{Isaac2016}. The nearest-neighbor distance $a$ is $0.256$~nm. This gives us a spin-diffusion constant of $D=Wa^2=15$~nm$^{2}$s$^{-1}$. This yields a diffusion length $l_D=2\sqrt{Dt}$. With a pulse length of $1$~s, we obtain a diffusion length of $7.6$~nm. Since this diffusion happens at both the inside and the outside of the resonant-slice, the actual expected broadening is approximately $15$~nm.

From the quality factor at each height combined with the piezovoltage we used to drive the cantilever, we estimate the drive amplitudes to be $60$, $37$ and $100$ nm, respectively, for the fitted thicknesses of $90$, $38$ and $31$ nm. Note that the cantilever's motion is in the $x$ direction, while in the calculation the resonant-slice thickness is in the radial direction. When the resonant-slice is broadened only in the $x$ direction, the spin signal is 3 times lower than when it is broadened in the radial direction. 

All the above effects cannot be easily summed, which makes it difficult to precisely determine which of the above processes is the limiting factor in the thickness of the resonant-slice. Altogether, we conclude that the NMR linewidth, and the motion of the cantilever together with spin diffusion qualitatively explain the experimentally found slice thicknesses. In most solids, $T_2$ will be larger than in copper. Therefore it is reasonable to assume that the resonant-slice width in these future experiments will be smaller. 

For the calculations of this experiment, we take the resonant-slice thickness to be uniform.

\section{Summary and outlook}

In conclusion, by performing nuclear magnetic-resonance force microscopy experiments down to $42$ mK, we demonstrate that the nuclear spin-lattice relaxation time $T_1$ can be detected by measuring the spin polarization in a volume of about $3000\times 3000\times 30$ nm$^3$; see Fig. \ref{fig:figure1}(c). A much smaller volume gives the same signal-to-noise ratio when a different sample than copper is measured and a smaller magnetic particle is used.  First,  copper is such a good conductor that our experiment is hampered by the eddy currents generated in the copper due to the motion of the magnetic-force sensor. The subsequent drop in quality factor increases the frequency noise in the measurements by an order of magnitude. This prevents us from performing experiments at smaller tip-sample heights. When the magnet is closer to the sample, the magnetization of the nuclei is also larger. Second, by improving the operation temperature of the fridge, future experiments can be performed at $10$ mK. 

By taking the ratio between the frequency noise within a bandwidth of $1$ Hz at a $Q>1000$ and the frequency shift of a single spin when the height between the magnet and sample is minimal, we find a volume sensitivity of $(30\ \text{nm})^3$.  Further improvements are possible when using smaller magnets on the force sensor \cite{Overweg2015}, which provide much higher field gradients. Previous measurements of the relaxation time with MRFM \cite{Alexson2012} have a volume sensitivity 3 orders of magnitude smaller \cite{Garner2004}. Our volume sensitivity is at least 4 orders of magnitude larger than what can theoretically be achieved in bulk NMR and even 10 orders larger than is usually reported in solid-state NMR \cite{Glover2002}. 

To substantiate our claim that using our technique new systems can be investigated, we discuss two examples. The first example is the measurements on nuclei that are in the vicinity of the surface state of a topological insulator. Bulk NMR on powdered samples show that the nuclei close to the surface state exhibit a Korringa-like relation \cite{Koumoulis2012}. Our technique could be used to directly measure inhomogeneities in the surface state, even if it dives below the surface. As a second example, we turn to superconducting interfaces between two oxides, such as the measurement of the two-dimensional electron gas between the interface of lanthanum aluminate (LaAlO$_3$) and strontium titanate (SrTiO$_3$) that becomes superconducting below $300$ mK \cite{Reyren2007}. The pairing symmetry of this superconducting state is still an open question, and by measuring T$_1$ of the nuclei in the vicinity of the interface it could be possible to obtain information about the pairing symmetry, as was done in 1989 with bulk NMR to show that the high Tc superconductor YBa$_2$Cu$_3$O$_7$  is an unconventional superconductor \cite{Hammel1989}. We believe that our technique to probe the electronic state through $T_1$ measurements with nanometer resolution will lead to new physics in the field of condensed matter at the ground state of strongly correlated electrons. 
 
\begin{acknowledgments}
We thank F. Schenkel, J. P. Koning, G. Koning and D. J. van der Zalm for technical support. This work is supported by the Dutch Foundation for Fundamental Research on Matter (FOM), by the Netherlands Organization for Scientific Research (NWO) through a VICI fellowship to T. H. O., and through the Nanofront program. T. M. K. acknowledges financial support from the Ministry of Science and Education of Russia under Contract No. 14.B25.31.0007 and from the European Research Council Advanced Grant No. 339306 (METIQUM).

J. J. T. W. and A. M. J. d. H. contributed equally.
\end{acknowledgments}

\bibliography{bibliography}
\end{document}